\documentclass[aps,prb,twocolumn,showpacs,superscriptaddress]{revtex4}
\usepackage{graphicx}
\usepackage{amssymb}
\usepackage{epstopdf}
\usepackage{amsmath}
\begin{document}
\title{Magnetic properties in layered (K, Rb and Cs)Co$_2$Se$_2$  with $ThCr_2Si_2-type$ structure}
\author{Jinhu Yang} \email{yang_jinhu@163.com}
\affiliation{Department of Physics, Hangzhou Normal University, Hangzhou 310036, China }
\affiliation{Department of Chemistry, Graduate  School of Science, Kyoto University, Kyoto  606 - 8502, Japan }
\author{Bin Chen}
\affiliation{Department of Physics, Hangzhou Normal University, Hangzhou 310036, China }
\affiliation{Department of Chemistry, Graduate  School of Science, Kyoto University, Kyoto  606 - 8502, Japan }
\author{Hangdong Wang}
\affiliation{Department of Physics, Hangzhou Normal University, Hangzhou 310036, China }
\affiliation{Department of Physics, Zhejiang University, Hangzhou  310027, China }
\author{Qianhui Mao}
\affiliation{Department of Physics, Zhejiang University, Hangzhou  310027, China }
\author{Masaki Imai}
\affiliation{Department of Chemistry, Graduate  School of Science, Kyoto University, Kyoto  606 - 8502, Japan }
\author{Kazuyoshi Yoshimura}\email{kyhv@kuchem.kyoto-u.ac.jp}
\affiliation{Department of Chemistry, Graduate  School of Science, Kyoto University, Kyoto  606 - 8502, Japan }
\author{Minghu Fang}\email{mhfang@zju.edu.cn}
\affiliation{Department of Physics, Zhejiang University, Hangzhou  310027, China }
\date{\today}
\begin{abstract}
 The magnetic properties of ThCr$_2$Si$_2$-type single crystals ACo$_2$Se$_2$ (A = K, Rb and Cs)  have been investigated by magnetic susceptibilities and isothermal magnetizations at various temperatures. The ferromagnetic phase transition temperatures are estimated as $\approx$ 74 K and 76 K and 62 K for A = K, Rb and Cs (in case of high magnetic field), respectively. The susceptibilities in the paramagnetic state obey the modified Curie-Weiss law quite well and the derived effective magnetic moments of the Co atom are about 2.21, 2.04 and 2.04 $\mu_B$/Co and the corresponding spontaneous moments derived at the ground state are 0.72, 0.59 and 0.52 $\mu_B$/Co as well as the generalized Rhodes-Wohlfarth ratios as 3.07, 3.42 and 3.96 for A= K, Rb and Cs, respectively. The magnetic moment aligns within ab-plane and a metamagnetism-like behavior occurs at a field of 3.5 T in CsCo$_2$Se$_2$ with $H$//ab-plane. The magnetic properties in this system were discussed within the frameworks of the self-consistent renormalization (SCR) and Takahashi's theory of spin fluctuations.
 \end{abstract}

\maketitle

\section{INTRODUCTION}
The layered compound AT$_2$X$_2$-type family shows versatile physical properties including antiferromagnetic (AFM) or ferromagnetic (FM) order,\cite{AFM} Fe-based superconductivity\cite{SC} and heavy fermion behavior.\cite{Steglich} The magnetic properties in this series of compound are very rich. For example, the parent of recent discovered Fe-As based superconductors AFe$_2$As$_2$ (A = K, Sr, Ba and Eu), \cite{Dongjk, Graser, Sutherland} is a collinear antiferromagnetic metal, while (Tl, K, Rb and Cs)Fe$_x$Se$_2$ with Fe-vacancies is a block AFM insulator and becomes a superconductor with superconducting temperature $\approx$ 30 K with decreasing Fe-vacancy levels.\cite{Fang} One of a significant common features of Fe-based superconductors is that the parent compounds show an antiferromagnetic phase transition at high temperature which can be suppressed by introducing holes or electrons into conducting layers (FeAs or FeSe). Furthermore, superconductivity also appears in the Ni analogues BaNi$_2$As$_2$\cite{Ronning}, SrNi$_2$As$_2$\cite{Bauer} and KNi$_2$Se$_2$,\cite{KNiSe} while  ferromagnetism or near the ferromagnetism in Co analogues KCo$_2$Se$_2$ \cite{Huan}, BaCo$_2$As$_2$\cite{Sefat}  were discovered.

Motivated by the versatility of spin fluctuations involved in AT$_2$X$_2$ structure, we turned out attention to the Co-based layered compounds. ACo$_2$X$_2$(A = K, Rb, Cs and Tl, X = S and Se ) has been investigated previously in polycrystalline by Greenblatt's group.\cite{Greenblatt, Huan, Huan2, Newmark} Their results shown that the distance of  intralayers or/and interlayers is of vital importance on the electrical and magnetic properties. In this structure the X-T-X (T: transition metals) forms layers, which are built up of edge-sharing TX$_4$ tetrahedra that extend two dimensionally in the ab-plane. For example, KCo$_2$Se$_2$ and KCo$_2$S$_2$ are ferromagnet with $T_C$ of around 80 and 120 K while CsCo$_2$Se$_2$ is antiferromagnetic order at low temperatures.\cite{Huan2,Oledzka} TlCo$_2$Se$_2$ is originally thought as a antiferromagnet with $T_N$ $\approx$ 90 K. \cite{Huan, Newmark} However, the later neutron results shown that  it is a non-collinear helical magnet with zero net spontaneous.\cite{Berger, Lizarraga} The Co atom are ferromagnetically arranged within the ab-plane while the magnetic moment of adjacent Co layers are rotated about 121$^o$ with respect to each other. \cite{Berger}

In this report, we investigated the temperature dependence of magnetic susceptibility and isothermal magnetization at various temperatures in high quality of single crystals $A$Co$_2$Se$_2$ (A = K, Rb and Cs), grown by self-flux method. All the samples show metallic luster and metallic conductivity and the ferromagnetic phase transition temperatures are estimated as $\approx$ 74 K and 76 K  for A = K and Rb, respectively. The susceptibilities obey the modified Curie-Weiss law in the paramagnetic state quite well. The derived effective magnetic moments of the Co atom from susceptibilities are about 2.21and 2.04 $\mu_B$/Co, and spontaneous moments at the ground state are derived as 0.72 and 0.59 $\mu_B$/Co. The generalized Rhodes-Wohlfarth ratios are estimated as 3.07 and 3.42 for A= K and  Rb, indicating itinerant ferromagnetism in this series. The magnetic moments align within ab-plane in this series. In case of Cs analogue, a metamagnetism-like occurs in a field of 3.5 T for $H$// ab-plane and the ferromagentic ordering are observed at high magnetic field below 62 K, contrasting to the previous report.\cite{Oledzka, Huan2} The spin fluctuation effect in this series were discussed within the frameworks of SCR and Takahashi theory of spin fluctuations.
\section{EXPERIMENTS}
Single crystals of ACo$_2$Se$_2$  (A =K, Rb and Cs) were grown by self-flux method. High purity starting materials K$_2$Se (obtained by the reaction of  K metal and Se powder at 573 K for 24 hrs) K (lump, 99.9\%), Co (powder, 99.99\%) and Se (powder 99.999\%) were used in the preparation of the single crystals. The stoichiometric amounts of starting elements  were  mixed carefully and sealed in a Al$_2$O$_3$ tube which was inserted into a quartz tube coated with carbon to avoid the possible reaction with the wall of the quartz tube with the starting materials. We first slowly heat the quartz tube to 673 K and dwell 5 hrs. Then the furnace  heated slowly up to 973 K (dwell 24h) and then heated up to 1323 K and keep at this temperature for 12 hours for homogeneity. Finally the furnace slowly cooled to 873 K at a rate of 3 K/h before the furnace was shut down. A typical size of the obtained single crystal  is of $\sim 2\times 3\times 1\times$mm$^3$. The  grown samples have golden luster and are very sensitive to the moisture. Therefore, all the above process are done in the Ar- filled glove box (H$_2$O $\leq$ 0.1 ppm and O$_2$ $\leq$ 0.1 ppm) and the samples are always kept in the glove box except for a very short time in the air during the preparation of magnetic measurements. The magnetic measurements were performed on Quantum Design Magnetic SQUID-VSM at Hangzhou Normal University.

\section{RESULTS AND DISSCUSSIONS}
\subsection{susceptibility}

\begin{figure}
\includegraphics[width=8cm]{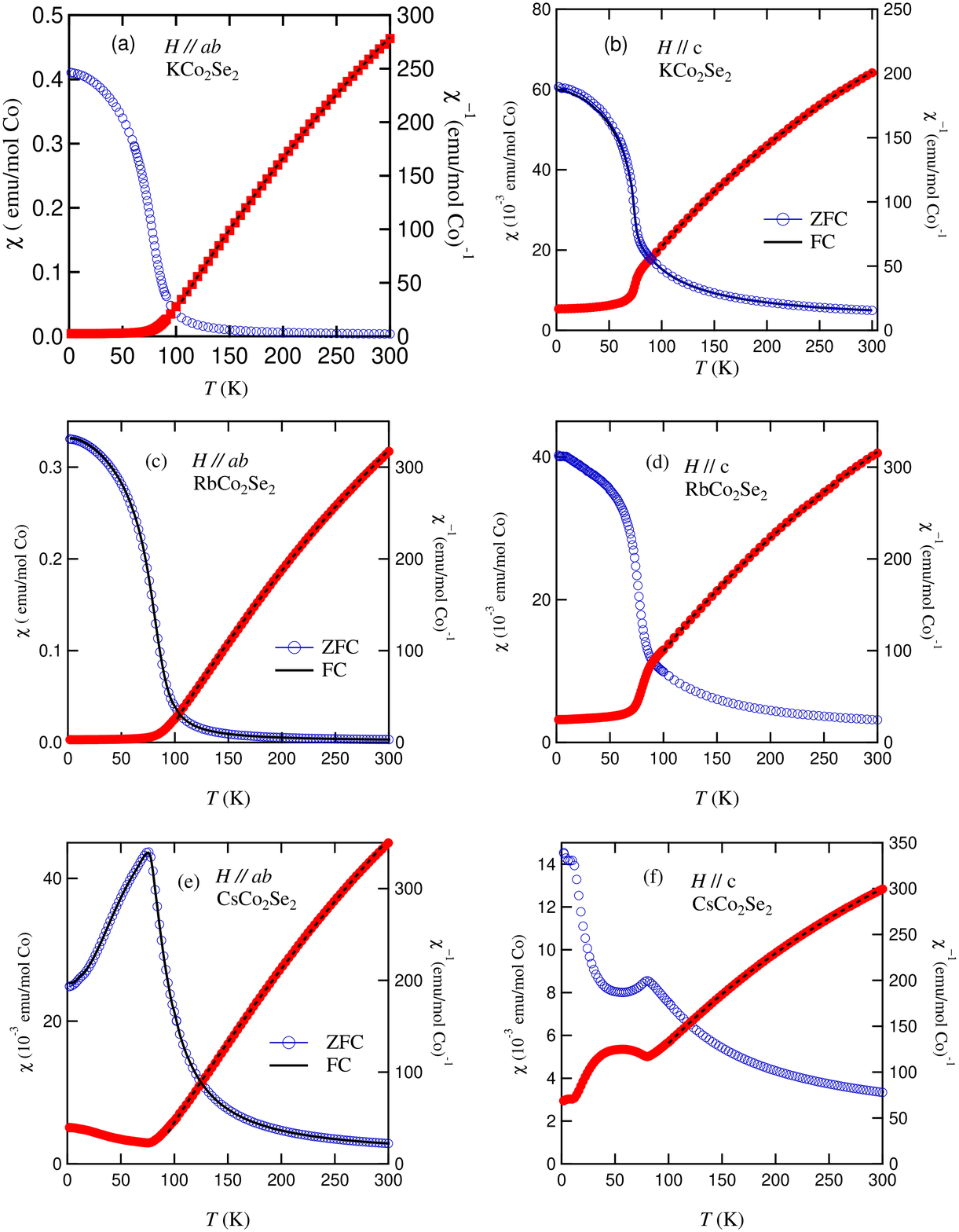}\\
\caption{(Color online) Temperature dependence of the magnetic susceptibilities $\chi(T)$=$M/H$ and reverse magnetic susceptibilities ($\chi^{-1}=H/M$) of the ACo$_2$Se$_2$ (A= K, Rb and Cs) under the magnetic field of 1 T parallel to ab-plane and $c$-axis, respectively. The dashed lines are the best fits by the modified Curie-Weiss law in high temperature region(100 K$\leq T \leq$ 300 K) and the solid line are the data obtained in field cooling (FC) process.}
\end {figure}

Figure 1 shows the temperature dependence of magnetic susceptibility ($\chi= M/H$) and reverse magnetic susceptibility ($\chi^{-1}=H/M$) for single crystals ACo$_2$Se$_2$ (A= K, Rb and Cs) in an applied magnetic field of 1 T parallel to ab-plane or c-axis, respectively.  Firstly, we discuss the $\chi$(T) for A=K and Rb single crystals. The quick increase of susceptibilities below at about 80 K when further decreasing the temperature is due to the occurrence of  a long range of ferromagnetic order and the susceptibilities tend to a saturated value of about 0.4 emu/mol Co at low temperatures. This saturated value  is only 15\% of that in polycrystal as previous reported,\cite{Huan} indicating the high quality of our single crystals. At high temperatures, $\chi$ can be well fitted by the modified Curie-Weiss law:
\begin{equation}
\chi=\chi_0+\frac{C}{T-\theta},
\end{equation}
where $\chi_0$ denotes the temperature-independent term, $C$ the Curie constant, and $\theta$ the paramagnetic Curie temperature. Therefore, one can derives parameters in the temperature from 100 K to 300 K for $H$//ab-plane or $H$// c-axis, respectively and the detail results are listed in Table I. The Curie constants are nearly independent of samples with the value of around 0.6 emu K/mol, with which one can estimates the effective magnetic moment $P_{eff}$ about 2.4 $\mu_B$ . This value is smaller than the theoretical value of 3.87 $\mu_B$ for a free Co$^{2+}$ ion, suggesting  the Co ion is of  itinerant character. The Curie temperature $\theta$ is almost independent of the samples for $H$//ab-plane or $H$//c-axis. However, the ratio of $\theta$ derived from the fitting for $H$//ab-plane to that of $H$//c-axis in single crystals A = K and Rb is about 2 and is increased to  about10 for CsCo$_2$Se$_2$ case. The positive $\theta$ means a ferromagnetic interaction between the adjacent interlayer Co moment or within the same layer is dominant. A large difference of $\theta$ derived from $H$//ab-plane and \textit{c}-axis indicates that the strength of interaction of Co atoms within the layer is stronger than it between the layers. On the other hand, the susceptibility in CsCo$_2$Se$_2$ displays a cusp-like peak around 80 K and seems due to an antiferromagnetic phase transition, which contrasts to the saturated behavior discovered in samples A = K and Rb. This series of samples A= K , Rb and Cs belong to soft ferromagnet due to no hysteresis  (data not shown) in  $M$ (H) curve and it is also indicated by the data measured both in  FC (field cooling) and ZFC (zero field cooling ) processes collapse into one curve.

\begin{table}[htbp]
\caption{Magnetic parameters of ACo$_2$Se$_2$ (A= K, Rb and Cs).$\chi_0$ $T_C$, $P_s$ and $P_{eff}$ are temperature-independent part of $\chi$(emu/mol Co), Curie constant C (10$
^{-3}$ emu/mol Co K), Curie temperature $\theta$ (K), spontaneous magnetization at ground state ($\mu_B$/Co atom) and effective moment ($\mu_B$/Co atom), respectively. }
\begin{center}
\begin{tabular}{lcccccccccc}
\hline\hline
sample & $\chi_0$& C& $\theta$&$P_{eff}$& $P_{c}$&$P_s$&$P_{eff}/P_s$  \\
 K$_{ab}$ &0.8& 0.61 & 85.2& 2.21 &1.41 &0.72& 3.07 \\
 K$_{c}$ &2& 0.84 & 38.2& 2.59 &1.78 &-& -  \\
 Rb$_{ab}$ &0.7& 0.52 & 88.5& 2.04 &1.27 &0.59& 3.46 \\
 Rb$_{c}$ &1& 0.57 & 37.4& 2.14 &1.36 &-&-  \\
 Cs$_{ab}$ &0.5& 0.52 & 75.8& 2.04 &1.27 &0.52&3.92\\
Cs$_{c}$ &1& 0.58 & 7.5& 2.15 &1.37 &-& - \\
  \hline\hline
\end{tabular}
\end{center}
\end{table}
\subsection{Isothermal Magnetization}

 \begin{figure}[tp]
\centering
\includegraphics[width= 8cm]{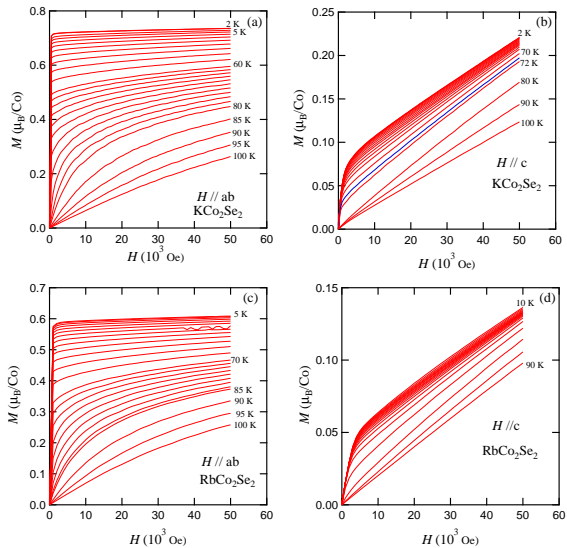}
\caption{ (Color online) Isothermal magnetization measurements for A= K and Rb single crystals at  temperatures from 2-300 K under $H//ab$-plane or $H//c$-axis, respectively. Not all data is shown. The steps of the temperature are 2 K or 5 K in the figure.  }
 \end {figure}
 
Figure 2 shows the results of magnetic  measurements for single crystals of KCo$_2$Se$_2$ and RbCo$_2$Se$_2$ in forms of isothermal magnetization $M$(H) curves and the corresponding Arrott plots at different temperatures  in  an applied magnetic field  $H//ab$-plane or $H//c$-axis, respectively. The magnetization goes quickly to a saturated magnetic moment 0.75 $\mu_B$  and 0.60 $\mu_B$  for KCo$_2$Se$_2$ and RbCo$_2$Se$_2$ at $H$//ab-plane, respectively. The magnetic moment increases with increasing magnetic field, however, there is no tendency to saturation for $H$//c-axis up to the field of 5 T. Figure 3 displays the magnetic field dependence of magnetic moment at various temperatures for  single crystal CsCo$_2$Se$_2$. A metamagnetism-like behavior occurs at the field 3 $\sim$ 4 T at field $H$//ab-plane and below the temperature of  around 64 K in $M(H)$ curve. The saturation moment is $\approx $ 0.55 $\mu_B$ /Co at the field of 5 T. No metamagnetism-like behavior were discovered for $H$// c-axis, but a slight kink were discovered in a similar field region at 10 K as shown in the inset of Fig.3. In the polycrystal  TlCo$_2$Se$_{2-x}$S$_x$ ($x$= 0.1.3 and 1.5), similar behavior was discovered in the field of 1 T which was ascribed to the possible competition between helix and a ferromagnetic phase at low temperature.\cite{Sabina}
   \begin{figure}[tp]
\centering
  \includegraphics[width= 8cm]{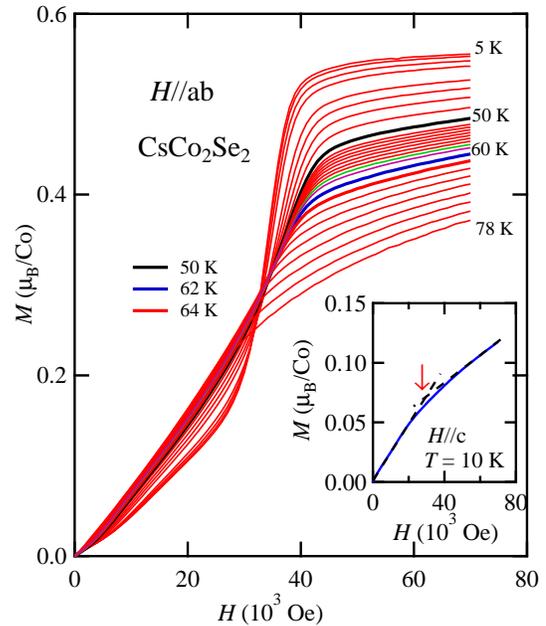}
    \caption{ (Color online) Isothermal magnetization measurements for A = Cs sample at various temperatures for $H$//ab-plane or $H$//c-axis. Inset: isothermal magnetization measurements for $H$//c-axis at 10 K. The steps of the selected temperature are 1 K, 2 K or 5 K in the figure and not all data is shown.  }
\end{figure}

According to Landau theory,  the free energy ($F$) can be expressed as the expansion of an order parameter. Therefore, from $H = \partial F/\partial M$, one can obtain $H = aM(T, H)+bM^3(T, H)$  or as often written as $ M^2(T, H) = M^2(T, 0) +B H/M(T, H)$ by neglecting the sixth and higher terms of the expansion, resulting so called Arrott plot. Figure 4 shows the Arrott plots for single crystals A = K and Rb for $H$//ab-plane or $H$//c-axis, respectively. A good linear behavior in the Arrott-plot of KCo$_2$Se$_2$  and RbCo$_2$Se$_2$ only exists at low temperatures and at high magnetic field side for $H$//ab-plane. In the Arrott-plot, one can estimates the spontaneous magnetic moment by extrapolating the linear relation to a positive intersection on y-axis. The spontaneous magnetic moment extrapolated from the linear relation in high magnetic region and at low temperatures to the y-axis is always positive and disappears above the temperatures which gives negative intersection to y-axis. It is clearly shown in Fig.4(a) and (c) that the spontaneous magnetic moments of A = K and Rb single crystals are aligned within ab-plane since the linear extrapolation to the y-axis for $H$//c-axis is negative at the lowest temperature of 2 K which means no spontaneous magnetic moment along this direction.
  \begin{figure}[tp]
\centering
  \includegraphics[width= 8cm]{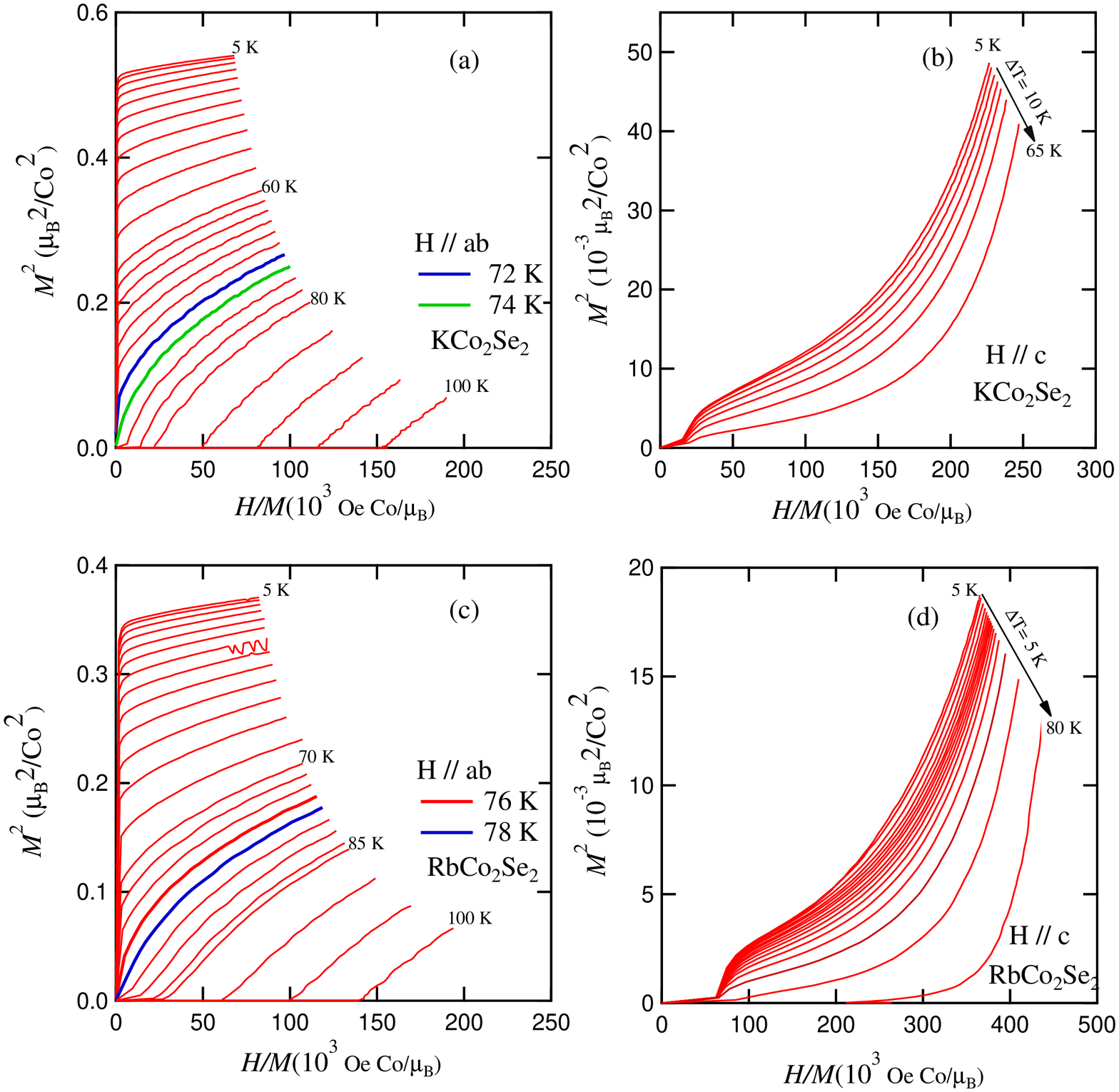}
\caption{(Color on line)  $M^2$ verse $H/M$  (Arrott plot)for A = K and Rb samples at various temperatures under $H$//ab-plane and $H$//c-axis, respectively. The steps of the selected temperature are 2 K or 5 K in the figure.}
\end{figure}
\begin{figure}[tp]
\centering
  \includegraphics[width= 8cm]{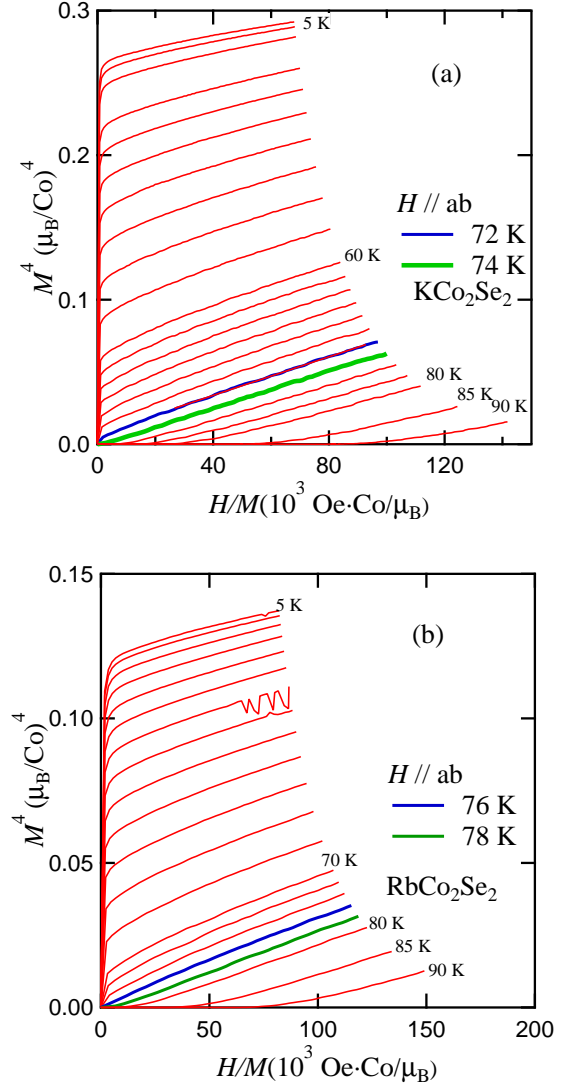}
\caption{ $M^4$ verse $H/M$ for A = K and Rb single crystal at various temperatures for $H$//ab-plane, respectively.  We measured isothermal magnetization from 2 to 300 K and only a part of data is shown here for clarity. The steps of the temperature are 2 K or 5 K in the figure.   }
\end{figure}
\begin{figure}[tp]
 \centering
  \includegraphics[width= 8cm]{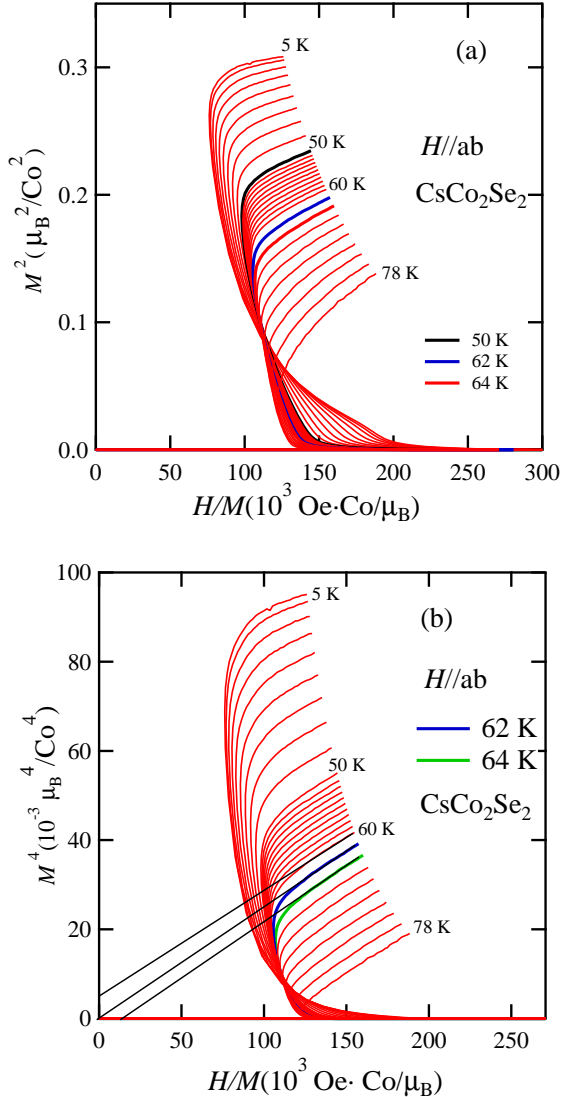}
\caption{(Color online) (a) $M^2$ verse $H/M$ and (b) $M^4$ verse $H/M$  at various temperatures for A = Cs sample for $H$//ab-plane. The steps of the selected temperature are 1 K, 2 K or 5 K in the figure. }
\end{figure}

 Generally,  $T_C$ can obtained by using Arrott plots since that the linear extrapolation passes through the origin at $T_C$. However, the curves show considerable deviation from linearity at high temperatures which makes it difficult to estimate the $T_C$ in this system. On the other hand, in SCR and Takahashi's theory for the itinerant-electron weak ferromagnet, $T_A$ and $T_0$ are the spectral widths of the dynamical spin-fluctuation spectrum and the dispersion of the dynamical magnetic susceptibility in wave-vector and energy space, respectively.\cite{SCR, Takahashi} These two important parameters dominate physical properties at ground state and finite temperatures. In Takahashi's theory the magnetization obeys the following relation at $T_C$ due to the disappearance of the fourth order expansion coefficient: \cite{Takahashi}
\begin{equation}
h=[T_A/3(2+\sqrt 5)T_C]^2p^5,
\end{equation}
\begin{table*}[t]
\caption{Spin-fluctuation parameters of ACo$_2$Se$_2$ (A = K, Rb and Cs). $T_C$ and $\bar{F}_{10}$ are derived from Arrott-plot. $T_0$ and $T_A$ are calculated according to SCR theory.  The $T_0^*$, $T_A^*$  are the best fits to the initial magnetic susceptibility data on Takahashi's theory. The estimated errors are shown in the brackets}
\begin{center}
\begin{tabular}{lcccccccccc}
\hline\hline
sample (A)& $T_C$ (K)& $T_0$ (10$^2$ K)& $T_A$ (10$^3$ K)&$\bar{F}_{10}$(10$^4$ K)&$\bar{F}_{10}^*$(10$^4$ K)&$T_0^*$ (10$^2$ K) & $T_A^{*}$(10$^3$ K)& \\\hline\hline
  K &74& 2.82 & 1.83& 3.15(0.01)&2.75(0.1)&3.06(0.1)&1.78(0.1)& \\
 Rb&76 &2.93&2.22&4.49(0.02)& 4.10(0.1) &4.23(0.1)&2.55(0.1)& \\
Cs &62 &3.48 & 2.58&5.10(0.02)&4.00(0.2)&4.02(0.1)&2.46(0.1)&\\
 \hline\hline
\end{tabular}
\end{center}
\end{table*}
Figure 5 shows $M^4$ dependence of $H/M$, indicating a good linear relation behavior in a wider range of temperature for crystals of A = K and Rb. Such behavior is also discovered in case of CsCo$_2$Se$_2 $ at high magnetic fields side shown in fig.6. Therefore, we estimated   $T_C$ in this series as 74 K ( K), 76 K (Rb) and 62 K (Cs) by linear extrapolation at  the temperature which passes the origins. The worse linearity of Arrott-plot suggests the existence of sixth-order nonlinear mode-mode coupling among spin fluctuation modes.\cite{SCR} In Takahashi's theory, the overall linearity of Arrott plot is due to a small value of $\eta =( T_C/T_0)^{1/3}$. As a result, very good linearities are observed experimentally in Arrott plot curves  in Ni$_3$Al($\eta$=0.25)\cite{yang}, Sc$_3$In($\eta$=0.22),\cite{Sc3In} while a relative larger value of $\eta$ gives worse linearities in Arrott plot curves as observed in MnSi ($\eta$=0.58),\cite{bloch} LaCoAsO($\eta \approx 0.46$)\cite{Ohta} and Fe$_3$GeTe$_2$($\eta \approx 0.46$).\cite{chen} We estimated $\eta$  for our single crystals A = K, Rb and Cs are 0.64, 0.64 and 0.57, respectively by using T$_C$ and T$_0$ listed in Table II below. The spontaneous magnetic moment $M_{0}(T)$ and the reciprocal initial magnetic susceptibility $1/\chi_{00}$ (= $\lim_{ H \to 0} H/M$)  are derived from the linear extrapolation in curves of $M^4$ vs $H/M$. From the estimated parameters above, one can obtained $P_{eff}/P_s $ and $T_C/T_0$. These values satisfy the so-called generalized Rhodes-Wohlfarth relation $P_{eff}/P_s = 1.4(T_C/T_0)^{-2/3}$ and Deguchi-Takahashi plot quite well as shown in Fig.7. \cite{RW, Takahashi}
\begin{figure}[tp]
\centering
  \includegraphics[width= 8cm]{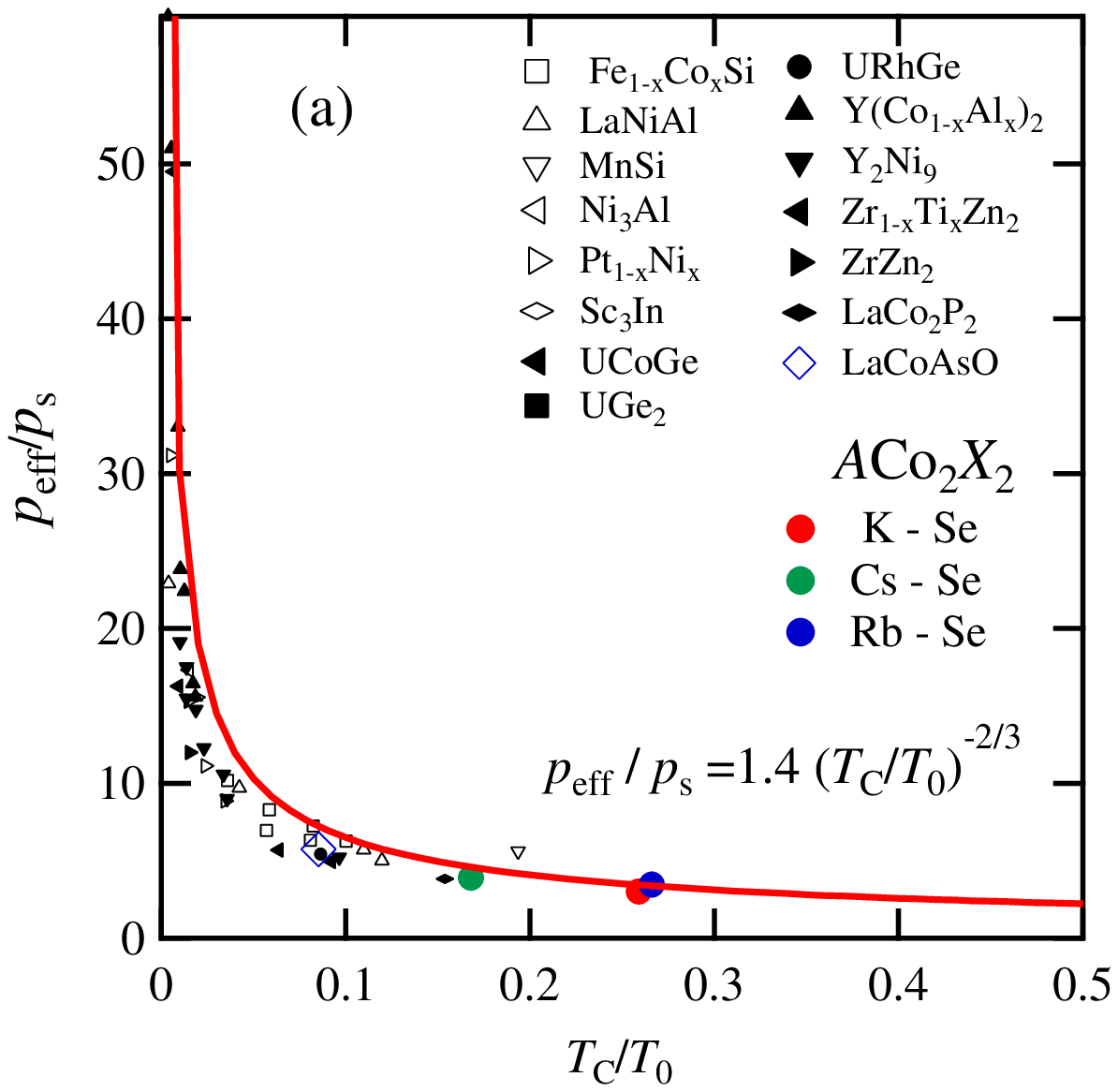}
   \includegraphics[width= 8cm]{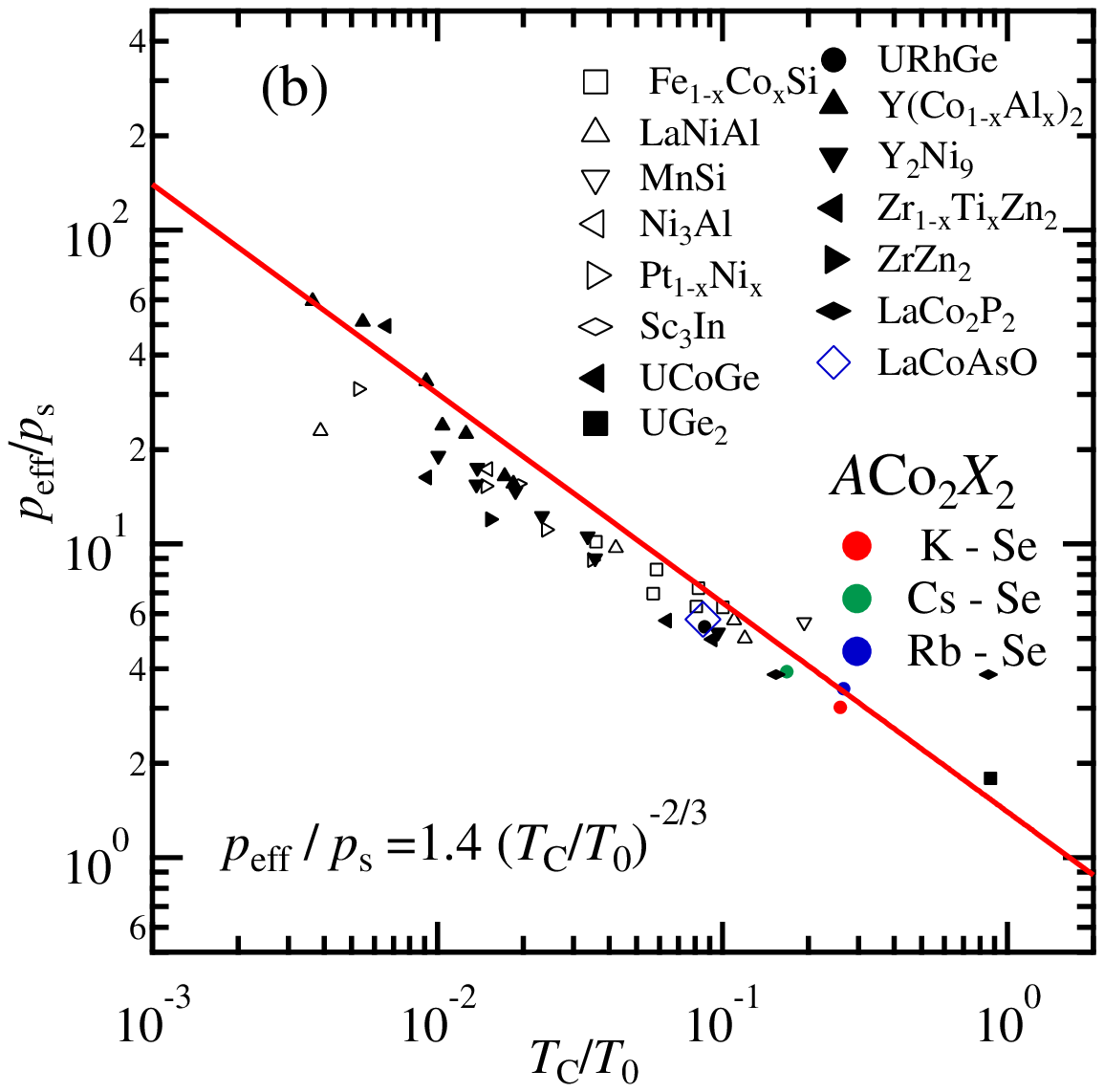}
\caption{(Color online) (a): Generalized Rhodes-wohlfarth plot. ACo$_2$Se$_2$ are plotted as closed circles.  (b): Deguchi- Takahashi plot. The other data are reproduced from Ref \cite{Ohta, SCR, MnSi, Ni3Al, Sc-In, ZrZn2, ZrTiZn2, FeCoSi, PtNi, YNi, YCoAl, LaNiAl} and references therein Ref (25). }
\end{figure}

Furthermore, another important spin-fluctuation parameter $\bar F_{10}$ also can be derived from Arrott-plot which is  the coefficient of $M^4$ term of the Landau expansion of free energy and usually expressed in Kelvin unit as
\begin{equation}
\bar F_{10}= N_A^3 (2\mu_B)^4/ \xi K_B
\end{equation}
where $N_A$ and $k_B$ are Avogadro's number and Boltzmann constant, respectively, and $ \xi$ is the slope of Arrott plot at $T_C$.  Therefore,  $\bar F_{10}$ can be determined uniquely from the slope of the Arrott-plot at $T_C$. Although the curves of the Arrott-plot ($M^2$ vs $H/M$) don't have a perfect linearity, it is reasonable to estimate the gradient at the temperatures of $T_C$ in a narrower high magnetic field region. We estimated $\bar F_{10}$ from a high field region of $M^2$ vs $H/M$ curve at $T_C$. In Takahashi's theory, $\bar F_{10}=\frac{4T_A^2}{15T_0}$ , and hence gives the relations below:
\begin{equation}
(\frac{T_C}{T_0})^{5/6}=\frac{\sqrt{30C_z}M_0(0)^2}{40C_{4/3}}(\frac{\bar F_{10}}{ T_C})^{1/2},
\end{equation}
\begin{equation}
(\frac{T_C}{T_A})^{5/3}=\frac{M_0(0)^2}{20C_{4/3}}(\frac{ 2T_C}{15C_z\bar F_{10}})^{1/3}.
\end{equation}
 where the values of the integral constants $C_{4/3}$ and $C_z$ are 1.006 and 0.50, respectively and $M_0(0)$ the spontaneous magnetic moment at ground state. The detailed spin fluctuation parameters are listed in Table II.

\begin{figure}[tp]
\centering
  \includegraphics[width= 8cm]{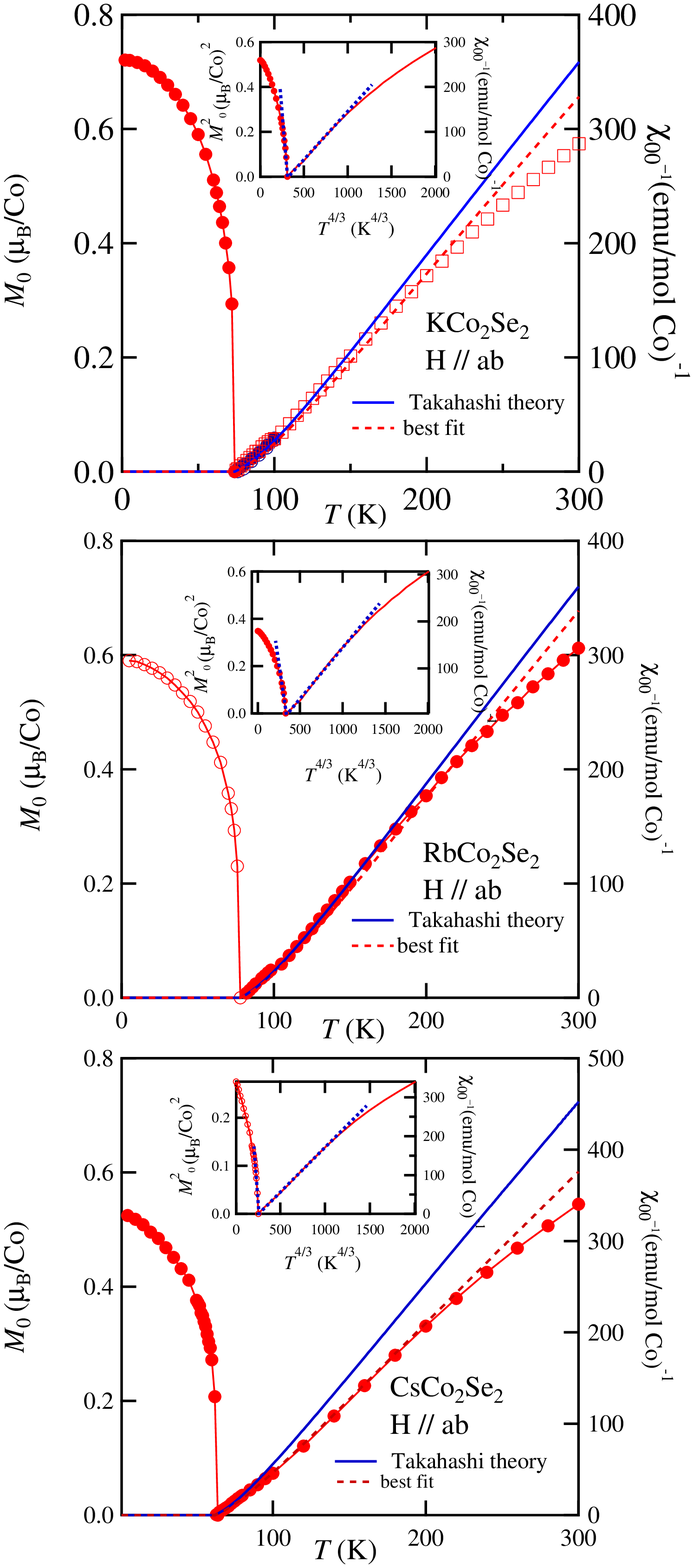}
\caption{(Color online)Temperature dependence of spontaneous magnetization ($M_0$  $Vs$ $T$) in ACo$_2$Se$_2$ (A= K, Rb and Cs)  and the initial susceptibilities for $H$//ab-plane. Insets: $M_0^2$ and $\chi_{00}^{-1}$ against $T^{4/3}$. }
\end{figure}

Using the spin fluctuation and thermal dynamic parameters, one can calculated the $1/\chi_{00}$ in a paramagnetic state by the SCR theory as
\begin{equation}
y \approx \bar f_1(-1+\frac{1+vy}{c}\int_{0}^{1/\eta}dzz^3[lnu-\frac{1}{2u}-\Psi(u)]),
\end{equation}
with $y^{-1} = 4\eta^2T_A\chi/3$, $u = z(y+z^2/t)$, $\eta = (T_C/T_0)^{1/3}$,$\bar f_1= F_{10}P_s^2/8T_A\eta^2$, $v = \eta^2T_A/U$, c = 0.3353, where $\Psi(u)$ is the digamma function,  $t = T/T_C$ and $U$ the intra-atomic exchange energy. Therefore the calculated temperature dependence of the inverse magnetic susceptibilities for samples of A = K, Rb and Cs by  using the parameters $T_0$, $T_A$, $\bar{F}_{10}$ and $M_0$ obtained from the experiment on Eq.(6) are shown in solid lines in Fig.8. The solid lines show clear deviations at high temperatures but agree well with the experimental data at low temperatures. If we allow the parameters $T_0$, $T_A$ and $\bar{F}_{10}$ to be changeable we can obtain the best fits to each sample as shown by dashed lines in the insets of Fig. 8. As a result, the parameters of  $T_0^*$, $T_A^*$ and $\bar{F}_{10}^*$ are a little larger than the values obtained from the Arrott-plots, however, in the same order of magnitude. In SCR and Takahashi's theory, the squared spontaneous moment $M_0^2$  and the initial susceptibility $\chi_{00}^{-1}$ have relation of $T^{4/3}$ near the $T_C$ as shown in the insets of Fig. 8. Very near $T_C$, the coefficients of $T^{4/3}$ is given as $a_c$ =1.27.\cite{Taka} We obtained the coefficients of $T^{4/3}$ for A= K, Rb and Cs are 1.27, 1.31 and 1.03, respectively by fitting the data near $T_C$. The results satisfied the theory very well except for the Cs case which is possible due to its very narrow $T^{4/3}$ temperature region. In Takahashi's theory, $T^{2}$-dependence of the squared spontaneous moment at low temperatures is described in terms of $T_A$ by neglecting the spin-wave contribution at low temperatures and obeys the formula: \cite{T2}
\begin{equation}
[\frac{M_0(0, T)}{M_0(0)}]^2=1-\frac{50.4}{M_0(0)^4} (\frac{T}{T_A})^2.
\end{equation}
 A good linearity is shown in curves of $[\frac{M_0(T)}{M_0(0)}]^2$ against T$^2$ at low temperatures as displayed in Fig.9. As results, the fitting by using formula (7) gives the values of $T_A$ are 1193, 1900 and 1828 K for samples of A = K, Rb and Cs, respectively. These values are very close to those derived from $\bar{F}_{10}$.
\begin{figure}[tp]
\centering
  \includegraphics[width= 8cm]{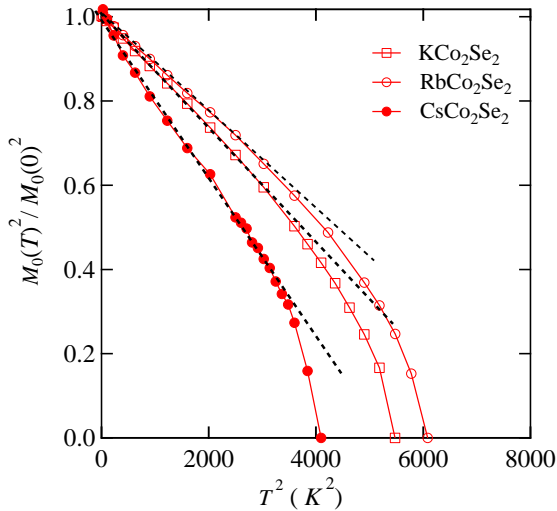}
\caption{(Color online)Temperature dependence of spontaneous magnetization in  single crystals ACo$_2$Se$_2$ (A= K, Rb and Cs). The dashed lines are the best fits to the data at low temperatures. }
\end{figure}
The parameters of $T_0$ and $T_A$ derived from the Arrott-plots or the initial magnetic susceptibility calculated on the SCR and Takahashi's theory are comparable with each other. Similar results were reported in LaCoAsO system. \cite{Ohta} It strongly indicates that the ferromagnetic spin fluctuations in Co analogues related to the new discovered Fe-based superconductors can be understood within the frameworks of SCR and Takahashi's theory. The study of spin fluctuations in ACo$_2$Se$_2$ (A= K, Rb and Cs) could give more information to shed light on the nature of superconductivity on Fe-based superconductors.

\section{CONCLUSION}
In summary, we investigated the temperature dependence of magnetic susceptibility and isothermal magnetization for the single crystals ACo$_2$Se$_2$ (A= K, Rb and Cs), which were grown by self-flux method. A ferromagnetic transition occurs at $\approx$ 74 K and 76 K for A= K and Rb. A metamagnetism-like occurs at about 3.5 T for $H$// c-axis in CsCo$_2$Se$_2$ below 62 K, and a cusp-like occurs at $\approx$ 80 K in susceptibility. The magnetic susceptibility obeys the modified Curie-Weiss law in the paramagnetic state quite well. The Arrott-plots show linear behavior at low temperatures and nonlinear behavior  in high temperatures. A good linear relation is realized in the curves of $M^4$ vs $H/M$ in a quite broaden temperature region, which is supported by a large $\eta$ in Takahashi's theory. The estimated spontaneous magnetic moments at zero temperature are 0.72, 0.58 and 0.52$\mu_B$ for A = K, Rb and Cs ( in high magnetic state), respectively. The according generalized Rhodes-Wohlfarth ratio are 3.07, 3.46 and 3.92, indicating a moderate itinerant ferromagnetism in this system. The spin fluctuation in this system can be understood in the frameworks of SCR and Takahashi's theory of spin fluctuations.

\section*{ACKNOWLEDGEMENTS}
This research is supported by the National Basic Research Program of China (973 Program) under grant No. 2011CBA00103, 2012CB821404 and 2009CB929104, the Nature Science Foundation of China (Grant No.10974175, 10934005 and 11204059) and Zhejiang Provincial Natural Science Foundation of China (Grant No. LQ12A04007), and the Fundamental Research Funds for the Central Universities of China (No. 2013FZA3003), and supported by Grant-in-Aid for the Global COE Program ``International Center for Integrated Research and Advanced Education in Materials Science" and by Grants-in-Aid for Scientific Research (19350030, 22350029) from the Japan Society for Promotion of Science.


\begin{thebibliography}{99}
\bibitem{AFM} M. Rotter, M. Tegel, D. Johrendt, I. Schellenberg, W. Hermes, and R. Pottgen, Phys. Rev. B \textbf{78},020503(R) (2008). J.-Q. Yan, A. Kreyssig, S. Nandi, N. Ni, S. L. Bud'ko, A. Kracher, R. J. McQueeney, R. W. McCallum, T. A. Lograsso, A. I. Goldman, and P. C. Canfield, Phys. Rev. B \textbf{78}, 024516 (2008). Z.Ren, Z. Zhu, S. Jiang, X. Xu, Q. Tao, C. Wang, C. Feng, G. Cao, and Z. Xu, Phys. Rev. B \textbf{78}, 052501 (2008).
\bibitem{SC} A. S. Sefat, R. Jin, M. A. McGuire, B. C. Sales, D. J. Singh, and D. Mandrus, Phys. Rev. Lett. \textbf{101}, 117004 (2008). A. Leithe-Jasper, W. Schnelle, C. Geibel, and H. Rosner, Phys. Rev. Lett. \textbf{101}, 207004 (2008). L. J. Li, Q. B. Wang, Y. K. Luo, H. Chen, Q. Tao, Y. K. Li, X. Lin, M. He, Z. W. Zhu, G. H. Cao, and Z. A. Xu, New J. Phys. \textbf{11}, 025008 (2009).
\bibitem{Steglich} F. Steglich, J. Aarts, C. D. Bredl, W. Lieke, D. Meschede, and W. Franz, and H. Schafer, Phys. Rev. Lett. \textbf{43}, 1892 (1979).
\bibitem{Dongjk} J. K. Dong, S. Y. Zhou, T. Y. Guan, H. Zhang, Y. F. Dai, X. Qiu, X. F. Wang, Y. He, X. H. Chen, and S. Y. Li, Phys. Rev. Lett. \textbf{104}, 087005 (2010).
\bibitem{Graser} S. Graser, A. F. Kemper, T. A. Maier, H. P. Cheng, P. J. Hirschfeld, and D. J. Scalapino, Phys. Rev. B \textbf{81}, 214503 (2010).
\bibitem{Sutherland} M. Sutherland, D. J. Hills, B. S. Tan, M. M. Altarawneh, N. Harrison, J. Gillett, E. C. T. OFarrell, T. M. Benseman, I. Kokanovic, P. Syers, J. R. Cooper, and S. E. Sebastian, Phys. Rev. B \textbf{84}, 180506 (2011).
\bibitem{Fang} Ming-Hu Fang, Hang-Dong wang, Chi-Heng Dong, Zu-Juan Li, Chun-Mu Feng, Jian Chen, and H. Q. Yuan, EPL \textbf{94}, 27009 (2011).
\bibitem{Ronning} F. Ronning, N. kurita, E. D. Bauer, B. L. Scott, T. Park, T. Klimczuk, R. Movshovich, and D. J. Thompson, J. Phys.:Condens. Matter \textbf{20}, 342203 (2008).
\bibitem{Bauer} E. D. Bauer, F. Ronning, B. L. Scott, and J. D. Thompson, Phys. Rev. B \textbf{78}, 172504 (2008).
\bibitem{KNiSe} J. R. Neilson, A. Llobet, A. V. Stier, L. Wu, J. Wen, J. Tao, Y. Zhu, Z. B. Tesanovic, N. P. Armitage, and T. M. McQueen, Phys. Rev. B \textbf{86}, 054512 (2012).
\bibitem{Huan} G. Huan, M. Greenblatt, J. Less-Common Met. \textbf{156}, 247 (1989), G. Huan, M. Greenblatt and K. V. Ramanujachary, Solid State Sommun. \textbf{71}, 221 (1989).
\bibitem{Sefat} A. S. Sefat, D. J. Singh, R. Jin, M. A. McGuire, B. C. Sales, and D. Mandrus, Phys. Rev. B \textbf{79}, 024512 (2009).
\bibitem{Greenblatt} G. Huan, M. Greenblatt and M. Croft, Eur. J. Solid state Inorg. Chem. \textbf{26}, 193 (1989).

\bibitem{Huan2}G. Huan, M. Greenblatt, and M. Croft, Eur. J. Solid State Inorg. Chem. \textbf{26}, 193 (1989).
\bibitem{Newmark} A. R. Newmark, G. Huan, M. Greenblatt, M. Croft, Solid State Commun. \textbf{71},1025 (1989).
\bibitem{Oledzka} M. Oledzka, J.-G. Lee, K. V. Ramanujachary, and M. Greenblatt, J. Solid Stata Chem. \textbf{127}, 151 (1996).
\bibitem{Berger} R. Berger, M. Fritzsche, A. Broddefalk, P. Nordblad, B. Malaman, J. Alloys compd. \textbf{343}, 186 (2002)
\bibitem{Lizarraga} R. Lizarraga, S. Ronneteg, R. Berger, A. Bergman, O. Eriksson, L. Nordstrom, Phys. Rev. B \textbf{70}, 024407 (2004).
\bibitem{Sabina} Sabina Ronneteg, Solveig Felton, Rolf Berger, Per Nordblad, J. Magn. Magn. Mater. \textbf{299}, 53 (2006).
\bibitem{SCR} T. Moriya, \textit{Spin Fluctuations in Itinerant Electron Magnetism} (Springer-Verlag, New York, 1985).
\bibitem{Takahashi} T. Takahashi, J. Phys. Soc. Jpn. 55, 3553 (1986).
\bibitem{yang} Jinhu Yang, Bin Chen, Hiroto Ohta, Chishiro Michioka, Kazuyoshi Yoshimura, Hangdong Wang, and Minghu Fang, Phys. Rev. B \textbf{83}, 134433 (2011).
\bibitem{Sc3In} B. T. Matthias, A. M. Clogston, H. J. Williams, E. Corenzwit, and R. C. Sherwood, Phys. Rev. Lett.\textbf{7}, 7(1961).
\bibitem{bloch} D. Bloch, J. Voiron, V. Jaccarino, and J. H. Wernick, Phys. Lett. \textbf{51A}, 259 (1975).
\bibitem{Ohta} H. Ohta, and K. Yoshimura, Phys. Rev. B \textbf{79}, 184407 (2009).
\bibitem{chen} in private communication.
\bibitem{RW}E. P. Wohlfarth, J. Magn. Magn. Mater. \textbf{7}, 113 (1978).
\bibitem{Taka} Y. Takahashi, J. Phys.: Condens. Matter \textbf{13}, 6323 (2001).
\bibitem{T2}T. Kanomata, T. Igarashi, H. Nishihara, K. Koyama, K. Watanabe,
K.-U. Neumann, and K. R. A. Ziebeck, Mater.Trans. \textbf{47}, 496 (2006), and reference therein.
\bibitem{MnSi}
D. Bloch, J. Voiron, V. Jaccarino and J.H. Wernick, Phys. Lett. A, \textbf{51}, 259 (1975).
\bibitem{Ni3Al} F. R. Deboer, C. J. Schinkel, J. Biesterbos and S. Proost, J. Appl. Phys. \textbf{40}, 1049 (1969).
\bibitem{Sc-In} J. Takeuchi and Y. Masuda, J. Phys. Soc. Jpn. \textbf{46}, 468 (1979).
\bibitem{ZrZn2} S. Ogawa, J. Phys. Soc. Jpn. \textbf{40}, 1007 (1976).
\bibitem{ZrTiZn2} S. Ogawa, J. Phys. Soc. Jpn. \textbf{25}, 109 (1968).
\bibitem{FeCoSi} K. Shimizu, H. Maruyama, H. Yamazaki and H. Watanabe, J. Phys. Soc. Jpn. \textbf{59}, 305 (1990).
\bibitem{PtNi} J. Beille, D. Bloch  and M. J. Besnus, J. Phys. F: Met. Phys. \textbf{4} 1275 (1974).
\bibitem{YNi} R. Nakabayashi, Y. Tazuke and S. Murayama, J. Phys. Soc. Jpn. \textbf{61}, 774 (1992).
\bibitem{YCoAl} K.Yoshimura, M. Takigawa, Y. Takahashi, H. Yasuoka, and Y. Nakamura, J. Phys. Soc. Jpn. \textbf{56}, 1138 (1987).
\bibitem{LaNiAl} A. Fujita, K. Fukamichi, H. Aruga-Katori and T. Goto, J. Phys.: Condens. Matter \textbf{7}, 401 (1995).
\end{thebibliography}
\end{document}